\documentclass[aps,twocolumn,showpacs]{revtex4}

\usepackage{epsfig}
\usepackage{amsfonts}
\usepackage{amssymb}
\usepackage{mathrsfs}
\usepackage{theorem}
\usepackage{amsmath}
\usepackage{times}
\usepackage{color}

\usepackage{ifpdf}
\ifpdf
\usepackage{epstopdf}   
\fi

\newtheorem{theorem}{Theorem}
\newtheorem{defin}{Definition}
\newtheorem{lem}{Lemma}

\newcommand{\ket}[1]{\ensuremath{\vert#1\rangle}}
\newcommand{\bra}[1]{\ensuremath{\langle #1\vert}}
\newcommand{\bk}[2]{\ensuremath{\langle #1\vert #2\rangle}}
\newcommand{\kb}[2]{\ensuremath{\vert #1 \rangle \langle #2 \vert}}

\newcommand{\sign}[0]{\ensuremath{\mathrm{sign}}}

\renewcommand{\vec}[1]{\ensuremath{\mathbf{#1}}}

\newcommand{\tr}{\ensuremath{\mathrm{tr}}}

\def\unity{\mbox{\small 1} \!\! \mbox{1}}

\def\unity{\mbox{\small 1} \!\! \mbox{1}}

\begin{document}
\title{Catalysis and activation of magic states in fault tolerant architectures}

\author{Earl T. Campbell}
\email{earltcampbell@gmail.com}
\affiliation{Department of Physics and Astronomy, University College London, Gower Street, London, WC1E 6BT, UK.}
\affiliation{Institute of Physics and Astronomy, University of Potsdam, 14476 Potsdam, Germany.}

\begin{abstract}
In many architectures for fault tolerant quantum computing universality is achieved by a combination of Clifford group unitary operators and preparation of suitable nonstabilizer states, the so-called magic states.  Universality is possible even for some fairly noisy nonstabilizer states, as distillation can convert many noisy copies into fewer purer magic states.  Here we propose novel protocols that exploit multiple species of magic states in surprising ways.  These protocols provide examples of previously unobserved phenomena that are analogous to catalysis and activation well known in entanglement theory.  
\end{abstract}


\maketitle  

Quantum computers are capable of executing algorithms whilst tolerating modest rates of faults or errors.   Stabilizer codes encode information in subspaces of larger Hilbert spaces and allow a proportion of errors to be actively detected and corrected~\cite{G01a}.  Whereas some \textit{anyonic} systems with topologically protected ground states provide a passive method of safely storing quantum information~\cite{Kit03}.  Research into anyonic systems has been stimulated by the recent discovery of alloys that are topological insulators~\cite{TopoInsulat08,Hsieh09}, opening up a variety of readily available systems that may be suitable for anyonic quantum computing. 

However,  fault tolerant quantum computing is not just about archiving quantum information, but also processing the information whilst stored in its protected form.  However, by employing stabilizer codes and topological systems we restrict how the quantum information may be fault-tolerantly manipulated.  Stabilizer codes only allow coherent implementation of a limited group of fault tolerant gates, the so-called \textit{transversal} gates.  Unfortunately, recent research has shown that no stabilizer code can both protect against generic errors and offer a universal set of transversal gates~\cite{Eastin09}. Similarly, topologically protected groups of gates, implemented by braiding anyons, are not universal for many species of anyons~\cite{Freedman06,Lachezar06,Bravyi06}.  Theoretically, some exotic anyons do offer universal topologically protected gates, but these are more physically speculative~\cite{Nayak08}.  Consequently, an alternative route to universal and fault tolerant quantum computing must be sought out.

This obstacle is overcome by gate injection techniques.  A suitable resource state is identified, and through fault tolerant gates and measurements, this resource is consumed in exchange for a new fault-tolerant unitary operator that promotes the group of gates to full universality.  For both stabilizer codes and anyonic systems, the manifestly fault tolerant gates are often contained within the Clifford group, the group of unitary operators that conjugate the Pauli operators.  What resource states might promote the Clifford group to universality?  Since the Clifford group maps stabilizer states --- eigenstates of Pauli operators --- to other stabilizer states, and such evolutions are efficiently classically simulable~\cite{AG01a}, we know that stabilizer states fail to provide universality.  However, numerous nonstabilizer states \textit{do} provide universality, including all single-qubit pure nonstabilizer states~\cite{Rei03a}.  Bravyi and Kitaev proposed the appellation \textit{magic} states for such resources~\cite{BraKit05}.  In their seminal article,  Bravyi and Kitaev showed that some mixed nonstabilizer states can enable universal quantum computing via a process of distillation into purer magic states.  Since preparation of the raw resources is not fault tolerant, we expect them to be noisy, and so distillation is essential.   

Some fault tolerance schemes actually provide a proper subgroup of the Clifford group, such as when braiding Ising anyons~\cite{Freedman06,Lachezar06,Bravyi06}.  Universality may still be possible via two levels of distillation if a resource state is available that first promotes the subgroup to the full Clifford group.  For example, Bravyi~\cite{Bravyi06} has shown that the aforementioned Ising anyon systems can be promoted to the full Clifford group by distilling certain noisy stabilizer resources.

The paradigm of magic states as a resource for promoting the Clifford group is analogous to other resource theories, such as: how entanglement is a resource when only local operations are available~\cite{HorodeckiReview}; and how continuous variable Gaussian entangled states can be utilized provided with just local Gaussian operations~\cite{Eisert03}.  In both these alternative examples of resource theories we have a thorough understanding of the fundamental principles behind what state transformations are possible.  The role of magic states is not yet understood as comprehensively as entanglement, although lately several results have begun to illuminate the subject.  Reichardt~\cite{Rei01a,Rei02a,Rei03a} provided several additional distillation protocols beyond those found by Bravyi and Kitaev.  He also identified some multi-qubit nonstabilizer states that can not, even probabilistically, be reduced to a single-qubit nonstabilizer state~\cite{Rei03a}.  Howard and van Dam~\cite{Howard09,Howard10a} studied the role of noisy unitary operators as resources.  They found that all depolarized single-qubit unitary operators that fall outside the Clifford group can enable universal quantum computing.  Campbell and Browne~\cite{Camp09c,Camp10a} identified an analog to bound entanglement, with certain families of nonstabilizer states being undistillable for finite-sized computers.  Ratanje and Virmani~\cite{Virmani10} considered resource theories that interpolate between separable states and stabilizer states and found new regimes that are efficiently classically simulable. 

This article explores the fundamental principles that govern magic states, and we uncover several new phenomena previously not observed.    Many of the new phenomena have analogous, though subtly distinct, counterparts in entanglement theory, such as entanglement catalysis~\cite{Catalysis99} and entanglement activation~\cite{HoroBound99}.  Previous work on magic states has focused on what is achievable with many copies of the same quantum state.  Whereas, a unifying theme of the protocols introduced here are the counterintuitive ways that two different sorts of resource can be jointly exploited.

Magic catalysis can be described as a scenario involving two agents: a ``magic state banker"; and an operator of a computer capable of only Clifford group operations.  The banker is willing to loan magic states to the operator, but requires that the operator returns \textit{exactly} the same quantum state at a later time.  We identify a protocol where the loaned magic state acts as a catalyst, enabling the operator to perform state transformations that would have been impossible otherwise.  Our protocol counteracts the misleading but intuitive idea that resources must be consumed to serve a function. 

Magic activation again involves a special resource, this time called the activator, that enables a probabilistic transformation that was impossible without this assistance.  This phenomena differs from catalysis in several key ways.   The activator is not returned to a banker, and the transformation may succeed with nonunit probability.  Furthermore, the probabilistic transformation also consumes a supply of bound magic states~\cite{Camp09c,Camp10a} that alone have limited computational power when in finite quantity. 

Next we discuss the existence of, the aforementioned, computationally weak multi-qubit states that were first identified by Reichardt~\cite{Rei03a}, which we call \textit{irreducible} non-stabilizer states.    The defining feature of irreducible non-stabilizer states is that, on their own, no single-qubit nonstabilizer state can be extracted from one copy.  We present new examples of irreducible non-stabilizer states for any number of qubits above two.  Next we introduce another new protocol that exploits a combination of irreducible non-stabilizer states and bound magic states.  Despite both resources being of limited utility we can, with some probability,  extract a magic state of arbitrarily high fidelity.  In many ways this protocol is more surprising than the previous magic state activation protocol.  However, this latter protocol relies on a large number of resources.  Depending on your preferred definition of activation, this protocol may also qualify as such.  However we prefer to stress its unique aspects and so refer to it as an \textit{asymptotic activation} protocol.

Combined, these results provide a significant step towards a complete understanding of the principles governing magic states and their manipulation.  Our results also prompt several interesting open problems that we discuss in the final section.

\section{Technical preamble}

In this section we refine our terminology and define notation, beginning with a quick review of stabilizer states and the Clifford group.   An $n$-qubit pure stabilizer state, $\ket{\psi}$, is a quantum state uniquely defined by $n$ commuting, and independent, Pauli operators $g_{j}$.  These operators generate by multiplication a group $\mathcal{S}$ of order $2^n$, the so-called stabilizer group for $\ket{\psi}$.  Every element of this group is said to stabilize the quantum state, such that $s \ket{\psi}=\ket{\psi}, \forall s \in \mathcal{S}$.  More generally, a mixed state is a stabilizer state if and only if it is an incoherent mixture of pure stabilizer states.  The Clifford group is the group of unitary operators that conjugate Pauli operators, such that for all Pauli operators $p$ we have $C p C^{\dagger}=p'$.  Equivalently, the Clifford group are the unitary operators that preserve the set of pure stabilizer states.  Important single-qubit Cifford unitary operators are the $H$ (Hadamard) and $T$ gates, which are best described in terms of their action on Pauli operators
\begin{eqnarray}
	H X H^{\dagger} = Z & ; & 	H Z H^{\dagger} = X ;  \nonumber \\
	T X T^{\dagger} = Y & ; & T Y T^{\dagger}= Z .
\end{eqnarray}
All single-qubit Clifford unitary operators can be decomposed into some sequence of these gates; that is, they generate the single-qubit Clifford group.  To generate the entire multi-qubit Clifford group we have to add an entangling gate, such as the well known control-not gate.  For further information on stabilizer states and the Clifford group, we refer the reader to Refs.~\cite{G01a,NC01b}.

Throughout we refer to a Clifford computer as follows.
\begin{defin}  A \textbf{Clifford computer} is a device capable of performing ideal Clifford unitary operators, preparation of stabilizer states, classical feedforward, classical randomness, Pauli measurements.
\end{defin}
For transformations implemented on such a device.
\begin{defin} If a Clifford computer can take an input state $\rho$ and deterministically output a state $\rho'$, then we denote this as $\rho \rightarrow_{D} \rho'$, and say that $\rho$ can be deterministically Clifford transformed to $\rho'$.  Conversely, if there exists no such Clifford transform, we denote this as $\rho \nrightarrow_{D} \rho'$.
\end{defin}
More generally, transformations may be probabilistic, as follows.
\begin{defin} If a Clifford computer can take an input state $\rho$ and with nonzero probability output a state $\rho'$, then we denote this as $\rho \rightarrow_{P} \rho'$, and say that $\rho$ can be probabilistically Clifford transformed to $\rho'$.  Conversely, if there exists no such probabilistic Clifford transform, we denote this as $\rho \nrightarrow_{P} \rho'$.
\end{defin}
The phenomena of catalysis and activation are essentially concerned with deterministic and probabilistic transformations respectively.  

The two most important single-qubit magic states are the eigenstates of the Clifford group unitary operators defined earlier, $H$ and $T$, such that
\begin{eqnarray}
\label{HTdefinitions}
	H \ket{H_{0}} = \ket{H_{0}} & ; &  	H \ket{H_{1}} = -\ket{H_{1}} ; \\ \nonumber
	T \ket{T_{0}} = e^{i \pi /3} \ket{T_{0}} & ; &  	T \ket{T_{1}} = e^{-i \pi /3} \ket{T_{1}} .
\end{eqnarray}
We also use similar notation for stabilizer states such as $Y$ eigenstates $\ket{Y_{0,1}}$.  For an $n$-qubit state a binary vector $\vec{v}=\{ v_{1},... v_{n} \}$ specifies the state 
\begin{equation}
	\ket{ H_{\vec{v}}} = \bigotimes_{j=1}^{n} \ket{H_{v_{j}}} ,
\end{equation}
and similarly for $\ket{T_{\vec{v}}}$.  Employing greek characters for mixed density matrices, we use
\begin{equation}
	\tau_{\vec{v}}  =  \kb{T_{\vec{v}}}{T_{\vec{v}} } ,
\end{equation}
with $\vec{v}$ again an $n$-bit vector.

\section{Magic catalysis}

\begin{figure}
\centering
\includegraphics{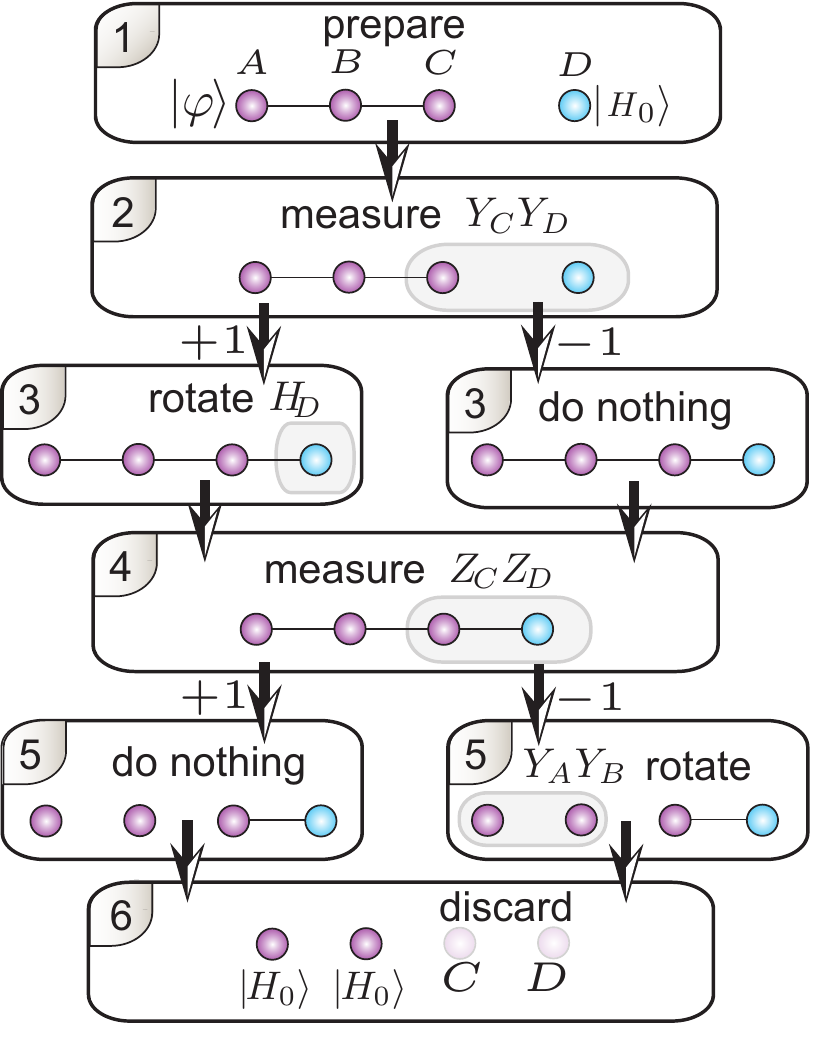}
\caption{An outline of the magic catalysis protocol.  Filled circles represent qubits, and lines between qubit denote correlations.  The protocol involves two measurements with random outcomes.  Although different measurement outcomes produce different projections, adaptively applied Clifford unitaries ensure the outcome is deterministic.  However, the determinism of our protocol relies heavily on the symmetries of the initial states.  The quantum states $\ket{\varphi}$ and $\ket{H_{0}}$ are defined in Thm.~\ref{thm:catalysis} and Eqn.~(\ref{HTdefinitions}), respectively.}
\label{fig:magic_catalysis}
\end{figure}

Here we present an example of magic catalysis.
\begin{theorem}
\label{thm:catalysis}
Magic catalysis is possible:  for the state $\ket{\varphi} \propto \ket{H_{0,0,0}}+\ket{H_{1,1,1}}$ we have $\ket{\varphi} \nrightarrow_{D} \ket{H_{0}}$ but with the addition of catalyst $\ket{H_{0}}$ we have $\ket{\varphi} \ket{H_{0}}\rightarrow_{D} \ket{H_{0}} \ket{H_{0}}$.
\end{theorem}
Clearly, this satisfies the constraints of the scenario described in the introduction since the process is deterministic and the catalyst is unchanged it can always be returned to the banker.   First we describe a protocol, also illustrated in Fig.~\ref{fig:magic_catalysis}, that implements the deterministic transformation  $\ket{\varphi} \ket{H_{0}}\rightarrow_{D} \ket{H_{0}} \ket{H_{0}}$. 
\begin{enumerate}
	\item Prepare the state $\ket{\varphi}$ on qubits $A,B,C$, and state $\ket{H_{0}}$ on qubit $D$;
	\item Measure the Pauli stabilizer $Y_{C}Y_{D}$;
	\item If the measurement yields outcome $+1$, then apply the unitary operator $H_{D}$;
	\item Measure the Pauli stabilizer $Z_{C}Z_{D}$;
	\item If the previous measure yields outcome $-1$, then apply the unitary operator $Y_{A}Y_{B}$;
	\item keep qubits $A$ and $B$, and discard qubits $C$ and $D$.
\end{enumerate}
Although the process involves two measurements with random outcomes, each measurement is conditionally followed by a unitary operator that ensures the same output regardless of the measurement outcome.  Consider step 3, after the $Y_{C}Y_{D}$ measurement with a $+1$ outcome, we have the state
\begin{eqnarray}
	H_{D} (\unity + Y_{C}Y_{D})\ket{\varphi}\ket{H_{0}} & = &  (\unity - Y_{C}Y_{D})\ket{\varphi} H_{D} \ket{H_{0}} , \nonumber \\ 
	& = &  (\unity - Y_{C}Y_{D})\ket{\varphi} \ket{H_{0}} ,
\end{eqnarray}
where the first line uses $H_{j}Y_{j}=-Y_{j}H_{j}$, and the second line uses $H\ket{H_{0}}=\ket{H_{0}}$.  Hence, we can deterministically implement a projection onto the $-Y_{C}Y_{D}$ subspace.  

Next, measurement results $-Z_{C}Z_{D}$  or $+Z_{C}Z_{D}$  give a projection of these qubits onto the state $\ket{\Psi^{-}}\propto \ket{1,0} - \ket{0,1}$ or $\ket{\Phi^{+}} \propto \ket{0,0}+\ket{1,1}$, respectively.  The use of $\ket{\Psi^{-}}$ projections plays a pivotal role throughout this article, effectively functioning as an odd parity projector for any basis.   That is, for any orthonormal basis $\{ \ket{b_{0}} , \ket{b_{1}} \}$ shared between two qubits we have $| \bk{\Psi^{-}}{b_{j}, b_{k}} | = (1-\delta_{j,k})/\sqrt{2}$, where $\delta_{j,k}$ is the Kronecker delta.  This feature of the singlet projector follows from $(U \otimes U)\ket{\Psi^{-}} \propto \ket{\Psi^{-}}$ for any unitary operator $U$, and so $\ket{\Psi^{-}}$ is odd-parity in any basis.   Returning to the problem at hand, the relevant basis is the Hadamard basis, where $\ket{\Psi^{-}}\propto \ket{H_{0,1}}- \ket{H_{1,0}}$.  Hence, the singlet projection picks out the second term of $\ket{H_{0,0,0,0}}+\ket{H_{1,1,1,0}}$, producing $\ket{H_{1,1}}\ket{\Psi^{-}}$, In accordance with step 5, we apply $Y_{A}Y_{B}$ (noting $Y$-gates flip Hadamard eigenstates) and discard the last two qubits.  This yields the desired output $\ket{H_{0,0}}$. 

If instead, step 4 provides a $+Z_{C}Z_{D}$ measurement outcome, we have a projection onto the state $\ket{\Phi^{+}}$, and so
\begin{eqnarray*}
	\bra{\Phi^{+}}_{C,D} \ket{\varphi}\ket{H_{0}} & \propto & \bra{\Psi^{-}}_{C,D} Y_{D} (\ket{H_{0,0,0,0}} + \ket{H_{1,1,1,0}}) , \\ \nonumber
	& \propto &  \bra{\Psi^{-}}_{C,D}  (\ket{H_{0,0,0,1}} + \ket{H_{1,1,1,1}}) , \\ \nonumber
	& \propto &   \ket{H_{0,0}} ,
\end{eqnarray*}
where the first line uses $\ket{\Phi^{+}} \propto Y_{D} \ket{\Psi^{-}}$, allowing further employment of the singlet projector.   Hence we yield the desired output regardless of measurement outcomes.  
  
To prove that we have identified a truly catalytic process, we must also show that the process was otherwise impossible, such that $\ket{\varphi} \nrightarrow_{D} \ket{H_{0}}$.  We actually proceed by showing the stronger result that $\ket{\varphi} \nrightarrow_{P} \ket{H_{0}}$, which directly entails the weaker deterministic no-go result.   Since we are attempting to probabilistically output a single-qubit state, we only have to consider Clifford transformations that project onto a stabilizer codespace, with a single logical qubit, and then decode~\cite{Camp09c}.  For a codespace with logical states $\ket{0_{L}}$ and $\ket{1_{L}}$, the result of projecting and decoding performs the transformation
\begin{equation}
	\ket{\varphi} \rightarrow \ket{\psi_{\mathrm{out}}} \propto \bk{0_{L}}{\varphi} \ket{0} + \bk{1_{L}}{\varphi} \ket{1}.
\end{equation}
For projections onto stabilizer subspaces, the ratio of the computational amplitudes,
\begin{equation}
\label{EQN:ratio}
	R( \ket{\psi_{\mathrm{out}}} ) = | \bk{0_{L}}{\varphi} |^{2}/| \bk{1_{L}}{\varphi} |^{2},
\end{equation} 
must be one of a few possible rational fractions~(see Appendix~\ref{APP:rational_fractions}).  However, for the target state, $\ket{H_{0}}$, this ratio is an irrational number $\tan^{2}(\pi/8)=3 - 2 \sqrt{2}$.  Hence, the exact transformation is impossible, and our proof is complete.  

The techniques in Appendix~\ref{APP:rational_fractions}  are sufficiently general to rule out many other Clifford transformations.  For example, if we ask whether $n$ copies of $\ket{\psi}$ can be exactly converted into a $H$-state, our method also proves this is impossible with finite $n$, and so $\ket{\psi}^{\otimes n} \nrightarrow_{P} \ket{H_{0}}$  (discussed further in Appendix~\ref{APP:many_copies_of_catalyst}).

In the preceding example of catalysis, both initial quantum states are pure.  However, there is a variant of the protocol where $\ket{\varphi}$ is replaced by a mixed state,
\begin{equation}
	\sigma_{\varphi} = \frac{1}{2} \left( \kb{\varphi}{\varphi} + H^{\otimes 3} \kb{\varphi}{\varphi} H^{\otimes 3} \right) .
\end{equation}
It is easy to prove that catalysis can be performed with $\sigma_{\varphi}$ because it is Clifford-equivalent to $\ket{\varphi}$; that is $\kb{\varphi}{\varphi} \rightarrow_{D} \sigma_{\varphi}$ and $\sigma_{\varphi} \rightarrow_{D} \kb{\varphi}{\varphi}$.  The first transformation is clearly possible by randomly applying $H^{\otimes 3}$ to $\ket{\varphi}$.  The converse transform is achieved by measuring the observable $Y_{A}Y_{B}Y_{C}$.  Given the $+1$ outcome we have the state $\ket{\varphi}$, whereas the $-1$ outcome produces $H^{\otimes 3} \ket{\varphi}$.  

Describing catalysis in terms of an interaction between a magic state banker and a computer operator gives it an operational flavor.  Although the scenario could be considered somewhat artificial.  We feel that the true depth of catalysis is that certain transformations become possible, for \textit{free}, assuming a reserve of magic-states.   Such transformations can be called \textit{magic-assisted} Clifford group operations, and it is an interesting open problem to determine the full structure of these operations.

\section{Magic activation}

Here we give an example of magic activation.  One of the distinguishing features of activation is that it utilizes resources from a family of bound states.  Rather than a general account magic-state boundness, for brevity we describe the concept with respect to noisy $T$-magic states,
\begin{equation}
\label{noisy_T_state}
	\tau(f) = f \tau_{0} + (1-f) \tau_{1} ,
\end{equation}
where $f$ is the fidelity and $f_{\mathrm{st}}=(1+1/\sqrt{3})/2$ is the threshold above which we have a nonstabilizer state.  The following statement follows directly from the more general results of Ref.~\cite{Camp10a}.
\begin{theorem}
\label{THM:activation}
	For any finite $n$, there exists a positive $\epsilon_{n}>0$, and a corresponding no-go region of fidelities $f \leq f_{\mathrm{st}} +\epsilon_{n} $.  Inside this no-go region, it follows that for any single qubit state, $\rho$, we have that $\tau(f)^{\otimes n} \rightarrow_{P} \rho$  if and only if $\tau(f) \rightarrow_{P} \rho$.  We say that the family of states $\tau(f)$ is bound.
\end{theorem}
Heuristically, this result instructs us that there exist nonstabilizer states where $n$ copies are no more useful than a single copy.  Since this holds even with probabilistic postselection, we cannot distill these states to higher purity.  There is clearly a parallel with bound entanglement, but there is also a subtle distinction.  The threshold fidelity, $f_{\mathrm{st}}+\epsilon_{n}$, below which the theorem applies, depends on the number of copies, $n$.  Hence, it is possible that the region shrinks as $n$ is increased,  maybe even such that $\epsilon_{n} \rightarrow 0$ as $n \rightarrow \infty$.  In contrast, bound entangled states are bound regardless of how many copies we have.   However, it is not known that the region actually does shrink.  Rather, it is merely a limitation of the techniques of Ref.~\cite{Camp10a} that this possibility has not been ruled out.   Whilst known techniques~\cite{BraKit05} can distill noisy $T$-states with fidelities greater than $(1+\sqrt{3/7})/2$,  there is no known method of distillation that functions below this fidelity.   Hence, it is possible that even for large $n$ the no-go region does not shrink below this level.  

\begin{figure}
\centering
\includegraphics{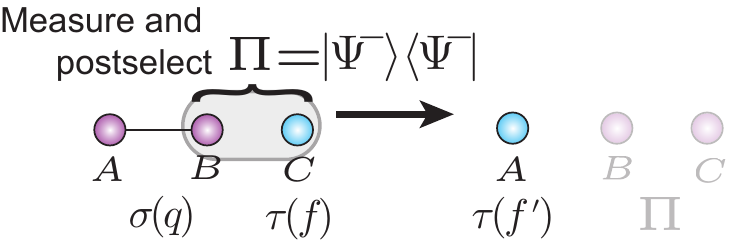}
\caption{An outline of the magic activation protocol which exploits the bound state $\tau(f)$ (see Eqn.~\ref{noisy_T_state}) in the presence of its activator $\sigma(q)$ (defined in Thm.~\ref{THM:activation}).  This single shot protocol succeeds probabilistically when qubits $B$ and $C$ are projected onto the singlet state.}
\label{fig:magic_activation}
\end{figure}

Subtleties aside, it is clear that $\tau(f)^{\otimes n}$, with sufficiently small fidelity and fixed $n$, cannot be distilled.  This is in contrast with noisy $H$-states, which are not a bound family of states.  For example, consider noisy $H$-states with \textit{any} initial fidelity large enough that no stabilizer decomposition exists.  With 7 copies of such noisy $H$-states one can implement a protocol~\cite{Rei02a,Rei03a} based on the \textsc{STEANE} code that, when successful, increases the fidelity~\footnote{Note that \textsc{STEANE} code distillation only reduces noise polynomially rather than exponentially, and so alone the protocol is not efficient.  However,  overall efficiency can be achieved by using this protocol to reach a threshold fidelity and then switching to another protocol.  For example, one may switch to implementing a protocol devised by  Bravyi and Kitaev~\cite{BraKit05} that utilizes 15 qubits per attempt.}.  The protocol must be iterated to achieve higher fidelities, and a unit fidelity is asymptotically approached with increasing $n$.  However, the important feature is that \textit{some} fidelity increase is always possible with finite copies, and that a similar protocol for all noisy $T$-states is ruled out by Thm.~\ref{THM:activation}.  This prompts the question, \textit{are very noisy $T$-states ever useful resources}?  We affirmatively answer this question by providing an activation protocol.
\begin{theorem}
\label{Thm_Activation}
Magic activation is possible:  for the activator $\sigma(q) = q \tau_{0,1}+ (1-q) \tau_{1,0} $ (for some $1>q>1/2$) and any $\tau(f)$, with $f_{\mathrm{st}}< f $, there exist a single-qubit state $\rho$ such that:
\begin{enumerate}
	\item[i.]  $\sigma(q) \otimes \tau(f) \rightarrow_{P} \rho$; even though 
	\item[ii.] $\sigma(q) \nrightarrow_{P} \rho$; and 
	\item[iii.]  $\tau(f) \nrightarrow_{P} \rho $.
 \end{enumerate}
\end{theorem}
Alone neither state can produce a particular output $\rho$, but combined it is possible.  The output state is again a noisy $T$-magic state, so $\rho=\tau(f')$.  Provided $f'>f$, condition (\textit{iii}) of the theorem immediately follows. 

We begin by describing the activation protocol, also illustrated in Fig.~\ref{fig:magic_activation}.
\begin{enumerate}
	\item prepare state $\sigma(q)$ on qubits $A,B$ and $\tau(f)$ on qubit $C$;
	\item measure the observables $Y_{B}Y_{C}$ and $Z_{B}Z_{C}$;
	\item postselect on $-1$ for both measurement outcomes, and discard qubits $B$ and $C$.
\end{enumerate}
The initial state can be expanded out as
\begin{eqnarray}
\label{eqn_initial_joint_system}
	\sigma(q)  \otimes \tau(f) & = & q f \tau_{0,1,0} + (1-q)f \tau_{1,0,0} \\ \nonumber
	 &  + & q(1-f) \tau_{0,1,1} + (1-q)(1-f) \tau_{1,0,1} .
\end{eqnarray}
The postselected measurements project qubits $B$ and $C$ onto the singlet state.  We use that $\ket{\Psi^{-}} \propto \ket{T_{1,0}}-\ket{T_{0,1}}$, and so
\begin{eqnarray}
	\bra{\Psi^{-}}  \sigma(q)  \otimes \tau(f)  \ket{\Psi^{-}}& \propto &q f \tau_{0} +  (1-q )(1-f) \tau_{1} .
\end{eqnarray}
We have effectively projected onto the odd parity terms of qubits $B$ and $C$, and then traced them out.  After renormalization, the state is $\tau(f')$ with fidelity
\begin{equation}   
	f' = \frac{q.f }{ q.f +(1-q)(1-f ) }
\end{equation}
It is easy to see that  $f' > f$ whenever $1>q > 1/2$, and so the transformation could not be achieved with $\tau(f)$ alone, satisfying condition (\textit{iii}).  To complete the proof we must show condition (\textit{ii}); that the transform could not be achieved with $\sigma ( q )$ alone.

The simplest transformation on $\sigma (q)$ alone is to measure qubit A of $\sigma (q)$ in the computational basis.  Due to the $T$-symmetry of the state, any Pauli basis gives the same result.   Hence for a $\pm 1$ outcome of any single qubit Pauli measurement, the resulting unnormalized state is
\begin{equation}
\label{eqn_unnormed_reduction}
\tr_{A}\left[ (1 \pm Z_{A}) \sigma(q) \right] \propto q c_{\pm}  \tau_{1} +  (1-q )(1- c_{\pm}) \tau_{0},
\end{equation}
where
\begin{equation}
	  c_{\pm}=\mathrm{trace} \left[  (\unity \pm Z_{A}) \tau_{0} \right]/2 = \left(  1 \pm 1/\sqrt{3} \right) /2 .
\end{equation}
Clearly, the ``$+1$" outcome gives a greater fidelity.  Furthermore, we have $c_{+}=f_{\mathrm{st}}$.  Renormalizing gives a noisy $T$-state with fidelity
\begin{equation} 
\label{1qubitReductFID}
  f'' = \frac{q.f_{\mathrm{st}} }{q.f_{\mathrm{st}}+(1-q)(1-f_{\mathrm{st}})}
\end{equation}
This fidelity fails to match that achieved by our activation protocol. However, a single qubit observable is clearly not the only option available, with many possible stabilizer measurements over both qubits.  Checking other possible measurements  (see Appendix~\ref{TwoQubitReductions}) one finds that the simple single-qubit measurement proves to be optimal.  Therefore, a single copy of $\sigma(q)$ cannot be probabilistically Clifford transformed to $\rho( f' )$,  the output of the protocol,  and so the activation is genuine.  Of course, our argument does not rule out that many copies of $\sigma(q)$ may accomplish this transformation, as is indeed the case (see Appendix~\ref{APP:Correlated_Noise_Resources}).  This feature is consistent with the analogous phenomena of activation in entanglement theory~\cite{HoroBound99}, as known entanglement activators are also many-copy distillable.

It is unclear whether more copies of the bound resource could be exploited to iterate or improve this particular magic state activation protocol.  However, the subsequent sections describe a more involved protocol that is stronger in two principle respects:  firstly it can be extended to consume arbitrarily many bound resources, with an output fidelity asymptotically approaching unity; secondly, the activating resources is also a computational weak state of a species that we introduce next.

\section{Irreducible non-stabilizer states}

\begin{figure}
\centering
\includegraphics{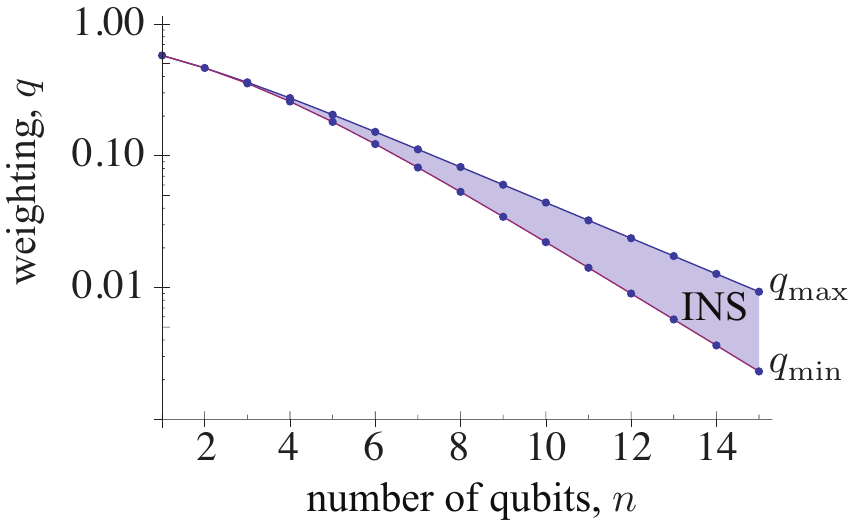}
\caption{A region of INS states of the form $\sigma_{\mathrm{INS}} (q,n)$, as in Eqn.~(\ref{eqn:MQD}).  The weighting, $q$, is shown on a log-scale, against the number of qubits $n$.  All states with the weighting satisfying $q_{\mathrm{min}}<q \leq q_{\mathrm{max}}$ are INS states, with the region being empty for $n=1,2$ and appearing for $n \geq 3$.  Some states outside the shaded region may also qualify as INS states.}
\label{fig:qminmax}
\end{figure}

This section introduces the notion of an irreducible non-stabilizer state, which is another form of noisy resource that is computationally weak.  We also present some new examples of such states to be used in the next section.
\begin{defin}
\label{def:IR}
A state $\sigma$ is an \textbf{irreducible non-stabilizer state} (an INS state) if both
\begin{enumerate}
	\item $\sigma$ is not a stabilizer state; and 
	\item  for all single-qubit nonstabilizer states, $\rho$, we have $\sigma \nrightarrow_{P} \rho$.
\end{enumerate}
\end{defin}
Reichardt identified the first examples of INS states~\cite{Rei03a}.  However,  Reichardt referred to them as \textit{counterexample} states, as he presented them to disprove a conjecture that all multi-qubit nonstabilizer states can be Clifford transformed to a single-qubit nonstabilizer state.  Obviously there are no single-qubit INS states, but Reichardt showed that two-qubit INS states do exist.  Despite being of limited computational power, some INS states prove useful when combined with other resources.  We consider states of the form
\begin{equation}
\label{eqn:MQD}
	\sigma_{\mathrm{INS}} (q,n) = q \tau_{0}^{\otimes n} + (1-q ) \unity / 2^{n} ;  
\end{equation}
which satisfy the definition of an INS state whenever the weighting, $q$, falls in a specific interval, $q_{\mathrm{min}} < q \leq q_{\mathrm{\mathrm{max}}}$, where
\begin{eqnarray}
		q_{\mathrm{max}} & = & [ 1 + (2 f_{\mathrm{st}})^{n-1}(\sqrt{3}-1)  ]^{-1} , \\
		q_{\mathrm{min}} & = & (2^{n}-1)/[(1+\sqrt{3})^{n}-1] .
\end{eqnarray}
Values of $q$ and $n$ satisfying these conditions are shown in Fig.~\ref{fig:qminmax}. 

First we show that for sufficiently pure states, we indeed have a nonstabilizer states, and so meet condition (1) of the definition.  In general, mapping out the space of multi-qubit mixed stabilizer states is an involved problem~\cite{Rei03a}.  However, there exists a simple witness that can detect many nonstabilizer states.  We introduce this witness in terms of a norm we call the stabilizer-norm (or just st-norm):

\begin{lem}
\label{lem:stab_norm}
A density matrix $\rho$, with decomposition in the Pauli basis $\rho=\sum_{j}a_{j} \sigma_{j}$, is a nonstabilizer state if
\begin{equation} 
 || \rho ||_{\mathrm{st}} = \sum_{j} | a_{j} | > 1 .
\end{equation}
\end{lem}
For single qubit states the condition is not just sufficient, but also necessary.  Indeed, for a single qubit state this inequality marks out an octahderon in the Bloch sphere.  However, there are many multi-qubit nonstabilizer states that are not detected by this witness.  To prove the lemma we first observe that the st-norm satisfies the triangle inequality, and hence is convex.  Furthermore, all pure stabilizer states, $\rho_{\mathrm{st}}$, have unit st-norm, $|| \rho_{\mathrm{st}} ||_{\mathrm{st}}=1$, and so no mixed stabilizer states can exceed unity.  

For the states of interest here,  the st-norm is
\begin{eqnarray}
\label{EQN:INS_state}
	|| \sigma_{\mathrm{INS}} (q, n )   ||_{\mathrm{st}} & = & q || \tau_{0}^{\otimes n} ||_{\mathrm{st}} + (1-q)/2^{n} , \\
	& = & q || \tau_{0} ||_{\mathrm{st}}^{n} + (1-q)/2^{n},
\end{eqnarray}
where the second line uses multiplicity of the st-norm with respect to the tensor product;  in general $|| \rho_{a} \otimes \rho_{b} ||_{\mathrm{st}}= || \rho_{a} ||_{\mathrm{st}} . || \rho_{b} ||_{\mathrm{st}}$.  Calculating $|| \tau_{0}||_{\mathrm{st}}=(1+\sqrt{3})/2$, and requiring the st-norm exceed unity, entails $q > q_{\mathrm{min}}$.  

Next we prove that for sufficiently impure states, condition (2) of our definition is satisfied.  It is well known~\cite{Camp09c} that such a transformation is impossible if it cannot be achieved by projecting onto a stabilizer codespace, with one logical qubit, and decoding.  First we note that all mixed single-qubit states with largest eigenvalue satisfying $\lambda \leq f_{\mathrm{st}}$ are stabilizer states.  We prove that, for $q<q_{\mathrm{max}}$, all codespace projections fail to achieve sufficient purity.  Hence, they output stabilizer states.  For a stabilizer code with projector $\Pi$, the projected state is
\begin{equation}
	\rho_{\mathrm{out}}  =  \frac{q.  \Pi \tau_{0}^{\otimes n} \Pi + (1-q) \Pi / 2^{n}}{q.  \tr ( \Pi \tau_{0}^{\otimes n}  ) + (1-q)/2^{n-1}  } .
\end{equation}
The largest eigenvalue of the projected state is
\begin{equation}
\label{EQN:lambda}
	\lambda = \frac{q. \tr (\Pi \tau_{0}^{\otimes n}  ) + (1-q)/2^{n} }{ q. \tr (\Pi \tau_{0}^{\otimes n}  ) + (1-q)/2^{n-1}}.
\end{equation}
To make further progress we must evaluate the maximum possible value of $ \tr (\Pi \tau_{0}^{\otimes n}  ) $.
\begin{lem}
\label{lem:graphcode}
For $n$ copies of a single-qubit state, $\tau_{0}$, and for all projectors, $\Pi$, onto a $2^{m}$-dimensional stabilizer subspace, the maximum probability of projection is
\begin{equation} 
 \max_{\Pi} \left[  \tr( \Pi  \tau_{0}^{\otimes n} ) \right] = f_{\mathrm{st}}^{n-m}
\end{equation}
\end{lem}
This lemma asserts that the maximum probability of any stabilizer projection is achieved by a series of single qubit stabilizer measurements.    The lemma can be proven using graph codes~\cite{GraphCodes1} as shown in appendix~\ref{App:graphcode}.  Applying the lemma (with $m=1$) to Eq.~(\ref{EQN:lambda})  gives a maximum achievable value of $\lambda$, which we denote with a star
\begin{equation}
	\lambda^{*}= \frac{q f_{\mathrm{st}}^{n-1} + (1-q)/2^{n} }{ q f_{\mathrm{st}}^{n-1}   + (1-q)/2^{n-1}} .
\end{equation}
If we wish to guarantee that the output is a stabilizer state, we require  $\lambda^{*} \leq f_{\mathrm{st}}$, and a little rearrangement produces the inequality $q \leq q_{\mathrm{max}}$.

Hence, we have proven the existence of a whole class of INS states using a very different approach to Reichardt.  Note that $q_{\mathrm{min}}$ and $q_{\mathrm{max}}$ differ only for three or more qubits, so our construction does not provide any two-qubit INS states.  

\section{Asymptotic magic activation}

Another new protocol is described here, which demonstrates two features not exhibited by the previous activation protocol.  Firstly, its exploits a combination of an INS state and many bound magic states.  Secondly, the output magic state can be arbitrarily pure,  asymptotically approaching unit fidelity with the number of bound states used.   In light of this, we distinguish this protocol by calling it \textit{asymptotic activation}.  For the purposes of this section, we consider the INS states with $q=q_{\max}$, as in Eq.~(\ref{EQN:INS_state}), and for brevity herein use the notation
\begin{equation}
\label{eqn:INS_qmax}
	\sigma_{\mathrm{INS}}(n)=\sigma_{\mathrm{INS}}(q_{\max},n) .
\end{equation}
Using this resource we have the following result.
\begin{theorem}
Asymptotic magic activation is possible:  for the INS state $\sigma_{\mathrm{INS}}(n)$ and any $\tau(f)^{\otimes n-1}$ (with $f>f_{\mathrm{st}}$), we have that $\sigma_{\mathrm{INS}}(n) \otimes \tau(f)^{\otimes n-1} \rightarrow_{P} \tau (f')$, where $f' \rightarrow 1$ as $n \rightarrow \infty$.
\end{theorem}

\begin{figure}
\centering
\includegraphics{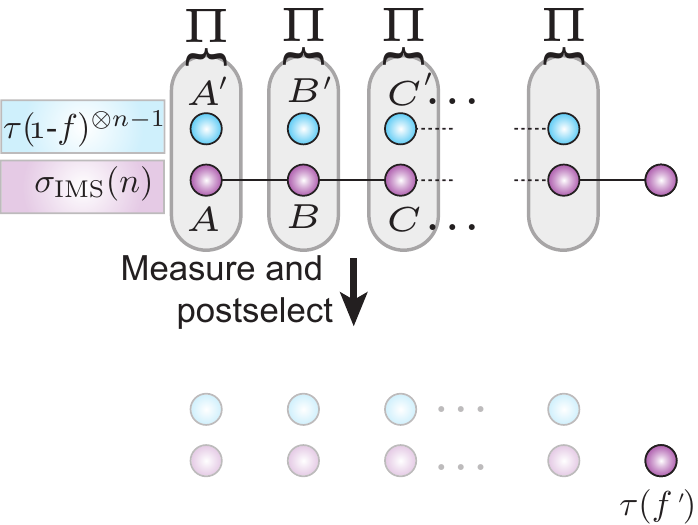}
\caption{An outline of the asymptotic magic activation protocol.  The protocol uses $n-1$ copies of the noisy $T$-states $\tau(f)$ (see Eqn.~\ref{noisy_T_state}) and a specific $n$-qubit activator $\sigma_{\mathrm{INS}}(q)$ (see Eqn.~\ref{eqn:INS_qmax} ).  The activator is an especially weak resource, known as an irreducible non-stabilizer state (defined in Def.~\ref{def:IR}).  The protocol pairs up each noisy $T$-state with a qubit from the activator, and succeeds when all $n-1$ pairs are projected onto the singlet state.}
\label{fig:magic_Asmyp_Activation}
\end{figure}

By definition, the INS state cannot be reduced to a single-qubit nonstabilizer state.  Since the bound states also resist distillation, the protocol seems to exploit some synergy between the two resources.  As with the previous activation protocol we utilize singlet projection.  The asymptotic activation protocol, also illustrated in Fig.~\ref{fig:magic_Asmyp_Activation}, is as follows.
\begin{enumerate}
	\item prepare $\sigma_{\mathrm{INS}}(n)$ on qubits $A, B, ....$, and prepare $\tau (f)^{\otimes n-1}$ on qubits $A', B',...$;
	\item flip every qubit of  $\tau (f)^{\otimes n-1}$ using the local Clifford $H.Y$ that maps $\tau_{0} \rightarrow \tau_{1}$;
	\item pair up $n-1$ qubits from each resource, pairing $A$ with $A'$ and $B$ with $B'$, etc;
	\item Measure the observables $X_{j}X_{j'}$ and $Z_{j}Z_{j'}$ for every pair;
	\item Postselect on ``-1" outcomes for every measurement outcome in every pair, and discard all measured qubits.
\end{enumerate}
After step 2, the quantum state is
\begin{eqnarray}
	\rho & = & q_{\mathrm{max}} \tau_{0}^{\otimes n} \otimes \tau(1-f)^{\otimes n-1} \\ \nonumber
	& & + (1-q_{\mathrm{max}}) \frac{\unity}{2^{n}} \otimes \tau(1-f)^{\otimes n-1}.
\end{eqnarray}
The subsequent steps project on the singlet state between paired up qubits, giving
\begin{equation}
	\rho \propto q_{\mathrm{max}} a^{n-1}   \tau_{0}  + (1-q_{\mathrm{max}}) b^{n-1} \tau_{1} ,
\end{equation}
where
\begin{eqnarray}
	a & = & \bra{\Psi^{-}} \tau_{0} \otimes  \tau(1-f) \ket{\Psi^{-}} = f/2 , \\ \nonumber
	b & = & \bra{\Psi^{-}} \unity \otimes \tau(1-f) \ket{\Psi^{-}}/2 = 1/4 .
\end{eqnarray}
Combining these equations and after some manipulation, we find that the output fidelity is
\begin{equation}
	f' = \left[ 1+ \left( \frac{f_{\mathrm{st}}} {f} \right)^{n-1}(\sqrt{3}-1) \right]^{-1} .
\end{equation}
Clearly this approaches unity in the large $n$ limit, provided that $f$ exceeds the stabilizer threshold.  

Unlike the previous activation protocol we are allowing for the number of copies to vary, with the phenomena becoming more pronounced in the large $n$ limit.  Since the number of copies is varying, and not fixed to some finite $n$,  previous no-go results on the non-distillability of $\tau(f)^{\otimes n}$ do not apply.  Consequently, we cannot guarantee that the transformation would be impossible without the addition of the INS state.  As we have not strictly proven $\tau(f)^{\otimes n} \nrightarrow_{P} \tau(f')$, we have exercised caution and not described this as a \textit{vanilla} activation protocol.  However, for small fidelities, $f < (1+\sqrt{3/7})/2 $, there is no known protocol~\cite{BraKit05} that performs this transformation, even in the limit of many copies.  The lesson this protocol teaches us is that large numbers of noisy $T$-states can be exploited to great effect when accompanied by another resource state.  The most fundamental open problem in this research area is now whether asymptotically many nonstabilizer states can be purified when completely unassisted by activating resources.

\section{Discussion and Conclusions}

We have introduced three protocols for quantum computers with Clifford-group unitary operators that are fault tolerant, and for clarity are taken to be ideal.   All our protocols make use of two different species of nonstabilizer states, which is a relatively unstudied topic compared with that of distilling many identical copies of a quantum state.  Each of the protocols is designed to illustrate a peculiar and counterintuitive phenomena that can occur in Clifford computers.   We now review each of these protocols and discuss related open problems.

Magic catalysis demonstrates that reserves of magic states do not have to be depleted to serve a function.   A magic state can act as a catalyst that enables a deterministic transformation that is impossible by Clifford transformations alone.   In our catalytic protocol, the catalyst was a Hadamard eigenstate, and the protocol depended on some very specific symmetries of this state.  This prompts the question of whether other nonstabilizer states can serve as catalysts.   For example, can eigenstates of the $T$-gate also act as catalysts?  We conjecture that ---  in light of deep underlying differences between $H$ and $T$-states --- the answer will be no.  The $T$ magic states are weaker in several regards.  Firstly, existing proposals for implementing non-Clifford gates do not directly exploit $T$-states.  Rather the $T$-states are probabilistically converted into states on the Bloch sphere equator (see Ref. \cite{BraKit05} or Appendix \ref{TwoQubitReductions}), and only then are they used for implementing a non-Clifford gate.  Secondly, noisy $T$-states just outside the set of stabilizer states are undistillable, or bound, in the sense reviewed earlier.  Beyond this anecdotal evidence, we have no firm proof that $T$-states cannot function as catalysts.  However, settling the conjecture either way should prove illuminating.

No protocol, prior to this article, has exploited noisy $T$-states arbitrarily close to the set of stabilizer states.  Indeed,  the evidence surveyed in the previous paragraph suggests that there exist noisy $T$-states, which despite being nonstabilizer states, cannot be utilized for any useful task.  However, our magic activation protocol shows that a noisy $T$-state combined with an activator resource can probabilistically output a single-qubit state that could not be achieved with either resource alone.  Hence, all noisy $T$-states outside the set of stabilizer state are useful for \textit{some} task.  Since all noisy $H$ nonstabilizer states are already known to be useful without the assistance of an activator, it is somewhat redundant to ask whether activation could be performed with $H$ states~\footnote{Although it is straightforward to check that a Hadamard analog of the activation protocol does work.}.  A dissimilarity with entanglement activation is that our magic activation protocol is not iterative, being defined for only a single round.  The most interesting questions on this topic concern what kinds of iteration are possible.  Our third and final protocol, asymptotic activation,  gives one possible extension.

Asymptotic activation shows that $(n-1)$ copies of any noisy $T$-state and a particular $n$-qubit resource can probabilistically output a magic state, which in the asymptotic limit approaches unit fidelity.  Furthermore, the special $n$-qubit resource is an irreducible non-stabilizer state, from which no single-qubit nonstabilizer states can be probabilistically extracted.  The class of irreducible non-stabilizer states is interesting in it own right, so our methods for constructing them may find applications elsewhere.  Indeed, one interesting problem is whether many copies of irreducible non-stabilizer states are distillable or a new form of bound state.  

Neither asymptotic activation nor standard activation are analogous to entanglement activation in every respect.  For example, in the entanglement activation of Ref.~\cite{HoroBound99} the protocol simultaneously exhibits the following three features:
\begin{enumerate}
	\item  The protocol can consume a variable number, $n-1$, copies of the bound resource with the output fidelity tending towards unity with increasing $n$;
	\item  The activating resource has a fixed size;
	\item  It is proven that neither the bound resources nor the activating resource can on their own be probabilistically transformed to the output of the activation protocol.
\end{enumerate}
Asymptotic activation has property (1), standard activation satisfies (2) and (3), but neither magic protocols simultaneously exhibit all three features.  This prompts the question, \textit{do magic protocols exist that are more sturdy analogs of entanglement activation with all three features}?  The extent of symmetries between the two resource theories is far from clear, and hence so is the answer to our posited problem.  

Considering all our protocols together, a key tool in all is the use of a singlet projection along with at least one state with multi-qubit correlations.  The singlet projector functions as a method of verifying if two qubits are nonidentical, although at the price of projecting those qubits into a stabilizer state.  Since our aim is to prepare nonstabilizer states, singlet projections can only be exploited when accompanied by multi-qubit correlations.  Indeed, we have seen that singlet projection is an extremely useful tool in this context.  So far we have not considered any scenarios with many copies of a multi-qubit correlated state, but it seems plausible that the singlet projection would prove useful in such contexts.  This is indeed the case, and for completeness we provide just such a strategy in Appendix~\ref{APP:Correlated_Noise_Resources}.  
 
\section{Acknowledgements}

We would like to thank Hussain Anwar for comments on the manuscript and conversations on this topic, and Dan Browne, Niel de Beaudrap and Matty Hoban for interesting discussions concerning magic states and Clifford computers.  This research was supported by the Royal Commission for the Exhibition of 1851 and the EU integrated project Q-ESSENCE.

\appendix

\section{}
\label{APP:rational_fractions}

Here we show that the ratio of amplitudes in Eq~(\ref{EQN:ratio}) can only take rational values, and hence cannot achieve the required irrational number.  Furthermore, the set of possible ratios is finite, so there is a limit to how closely the target ratio can be approximated.   We present a very general form of the proof, which can be used to rule out many other Clifford transformations.  The techniques introduced here indicate that there may well be hope for building a general framework for understanding catalysis.  

Before beginning the core proof, we shall make some observations;  The specific initial state of interest, $\ket{\varphi}$, is Clifford equivalent to
\begin{equation}
\label{eqn_evenTerms}
	 \ket{\varphi'}  = ( \ket{0,0,0} + i \ket{0,1,1}+ i \ket{1,0,1}+ i \ket{1,1,0} ) / 2.
\end{equation}
The required Clifford is simply $\sqrt{X}^{\otimes 3}$, which maps the Hadamard eigenstates to the Bloch sphere equator such that:
\begin{equation}
	\sqrt{X} \ket{H_{j}} = \ket{H'_{j}} = ( \ket{0} +(-1)^j e^{i \pi /4} \ket{1})/ \sqrt{2}.
\end{equation}
Hence, an equal superposition of $\ket{H'_{0,0,0}}$ and $\ket{H'_{1,1,1}}$ cancels out odd excitation terms, leaving only the even terms shown in Eqn.~(\ref{eqn_evenTerms}).  Next we recall that, stabilizer states must have the form~\cite{Dehaene03}
\begin{equation}
\label{eqn:stab_normal_form}
	\ket{\psi_{\mathrm{st}}} = \sum_{\vec{x} \in C+\vec{y}} i^{l(\vec{x})}(-1)^{q(\vec{x})} \ket{\vec{x+y}} / \sqrt{|C|} ,
\end{equation}
where $l(\vec{x})$ and $q(\vec{x})$ are some linear and quadratic functions in $\vec{x}$, $C$ is a binary linear subspace and $\vec{y}$ is some constant binary vector.   Notice that our state $\ket{\varphi'}$ has a very similar form to stabilizer states as in the computational basis the coefficient have equal magnitude and phases are multiplies of $i$.  Such states are interesting and deserving of their own title.
\begin{defin} We say a pure quantum state, $\ket{\psi}$, is a \textbf{pseudo-stabilizer} state if and only if there exists a Clifford unitary $U$ such that:
\begin{equation}
\label{eqn:pseudo_normal_form}
	U  \ket{\psi} = (\sum_{\vec{x} \in \mathcal{P}}  i^{f(\vec{x})} \ket{\vec{x}})/ \sqrt{\mathcal{|P|}},
\end{equation}
where $\mathcal{P}$ is a set of $n$-qubit bit strings, and $f: \vec{x}\rightarrow \{0,1\}$.  Furthermore, we say a pseudo-stabilizer state has complexity $P$, such that
\begin{equation}
	P ( \ket{\psi}  )  = \mathrm{min} \left\{ | \mathcal{P}| \bigg| U \ket{\psi}  = \sum_{\vec{x} \in \mathcal{P}} \frac{ i^{f(\vec{x})} \ket{\vec{x}} }{ \sqrt{\mathcal{|P|}}} ; \forall U \in \mathcal{C}\right\} .
\end{equation}
This is simply the smallest possible $|\mathcal{P}|$ over all valid decompositions.
\end{defin}
Notice that genuine stabilizer states also satisfy this definition but have trivial complexity $P=1$, and the decomposition of Eqn.~(\ref{eqn_evenTerms}) entails that $P( \ket{\varphi'} ) \leq 4$.  In contrast the $H$-states are not pseudo-stabilizer states as defined above.

Here we prove that the complexity of a pseudo-stabilizer state limits the possible single-qubit states one can produce by Clifford transformations.
\begin{theorem}
\label{eqn:catalysis_no_go}
	Consider a pseudo-stabilizer state $\ket{\psi}$ of complexity $P$.   If $\ket{\psi}\rightarrow_{P} \ket{\psi_{\mathrm{out}}}$ where $\ket{\psi_{\mathrm{out}}}$ is a pure single qubit state, then it follows that the amplitude ratio satisfies
	\begin{equation}
		R( \ket{\psi_{\mathrm{out}}} ) = \frac{ |\bk{0}{ \psi_{\mathrm{out}}} |^{2}}{|\bk{1}{\psi_{\mathrm{out}}} |^{2}} \in R_{p} ,
	\end{equation}
	where $R_{P}$ is the set of feasible ratios
	\begin{equation}
	 	R_{P} = \left\{  \frac{ a_{0}^{2}+b_{0}^{2}}{a_{1}^{2}+b_{1}^{2}}  \bigg| a_{j},b_{j} \in \mathbb{Z} ; |a_{j}|+|b_{j}| \leq P   \right\} .
	\end{equation}
	Conversely, for any $\ket{\psi'} $ with $R( \ket{\psi'} ) \notin R_{P}$ we can conclude $\ket{\psi}\nrightarrow_{P} \ket{\psi'}$.
\end{theorem}
Notice that all of the feasible ratios from a pseudo-stabilizer state are rational fractions, so exactly producing a $H$-state is impossible.  The specific result $ \ket{\varphi} \nrightarrow_{P} \ket{H_{0}} $ then follows from our earlier observation that $\ket{\varphi}$ is a pseudo-stabilizer state of bounded complexity $P(\ket{\varphi})\leq4$.

As noted in the main text, we only have to consider probabilistic Clifford transforms that project onto a single qubit stabilizer subspace and then decode, and so we can achieve
\begin{equation}
	R( \ket{\psi_{\mathrm{out}}} ) = | \bk{0_{L}}{\psi} |^{2}/  | \bk{1_{L}}{\psi} |^{2} ,
\end{equation}
where $\ket{0_{L}}$ and $\ket{1_{L}}$ are logical states of the stabilizer subspace, which by Eqn.~(\ref{eqn:stab_normal_form}) can be expressed as
\begin{equation}
\label{eqn:logical_ops}
	\ket{0,1_{L}} = \sum_{\vec{x} \in C_{0,1}+\vec{y_{0,1}}} \frac{ i^{l_{0,1}(\vec{x})}(-1)^{q_{0,1}(\vec{x})} \ket{\vec{x}}}{ \sqrt{|C_{0,1}|}} ,
\end{equation}
with the numeric subscripts differentiating $(C, \vec{y}, q, l)$ for the two states.  It is well known that logical states of stabilizer codes can always be found such that they differ by Pauli rotations, such that $\ket{1_{L}}=X_{L}\ket{0_{L}}$.  Pauli operators can change $\vec{y}$, $l$ and $q$, but not $C$ and so $C_{0}=C_{1}=C$.

Using equation~(\ref{eqn:pseudo_normal_form}) and~(\ref{eqn:logical_ops}) we find that:
\begin{equation}
	 \bk{0,1_{L}}{\psi'} =  \sum_{\vec{x} \in \mathcal{P} \cap ( C + \vec{y_{0,1}}) }  \frac{i^{l_{0,1}(\vec{x})+f(\vec{x})}(-1)^{q_{0,1}(\vec{x})} }{ \sqrt{|C|.|\mathcal{P}|}},
\end{equation}
each term in the summation is a multiple of $i$ and there are no more than $P(\ket{\psi'})=| \mathcal{P} |$ terms.  Hence we have
\begin{equation}
	 \bk{0,1_{L}}{\psi'} =  (a_{0,1}+i b_{0,1})/ \sqrt{|C|.|\mathcal{P}|} ,
\end{equation}
where $a_{j}, b_{j} \in \mathbb{Z}$ and the limited number of terms entails $|a_{j}|+|b_{j}|\leq P(\ket{\psi'})$.  Calculating the ratio of these amplitudes, the $|C|.|\mathcal{P}|$ factors cancel and we have the result as stated in Thm.~\ref{eqn:catalysis_no_go}.

From an infinite set of rational numbers, one can always find an arbitrarily good approximation to any irrational.  However, the set of feasible ratios is limited by the constraints $|a_{j}|+|b_{j}| \leq P(\ket{\psi})$ and and so the set of possibilities is not just finite but potentially very small.  We have presented an argument based on rationality for generality.  However, it is quite straightforward to numerically search the limited set of possibilities and verify that for $P(\ket{\varphi})=4$ we can never achieve $R=\tan (\pi /8)^{2}$.  Such a search produces $1/5$ as the closest possibility, which differs from the target by over $0.028$.  Note that Thm.~(\ref{eqn:catalysis_no_go}) places a restriction on feasible ratios, but does not guarantee that all such ratios are achievable.

The theorem deduced here rules out many Clifford transforms, but is far from the generality of the majorization criteria that is used in entanglement theory~\cite{Catalysis99}.  In entanglement theory, the majorization criteria depend on the coefficients of the quantum state in the Schmidt basis.  If we consider all possible \textit{local} unitaries, and rotate a state to have the minimal possible support in the computational basis then this also yields the all important Schmidt coefficients.  Returning to the context of magic states, our approach hints that minimizing support over all possible \textit{Clifford} unitaries also gives a decomposition with important coefficients.   Our investigations into this approach are ongoing. 

\section{}
\label{APP:many_copies_of_catalyst}

Here we briefly address the question of whether many copies of $\ket{\varphi}$ can be probabilistically Clifford transformed into $\ket{H_{0}}$.  Much of the technical apparatus required was established in Appendix~\ref{APP:rational_fractions}.  Given that $\ket{\varphi}$ is a pseudo-stabilizer state with complexity $P(\ket{\varphi}) \leq 4$, it follows that $\ket{\varphi}^{\otimes n}$ is also a pseudo-stabilizer state but with $P(\ket{\psi}) \leq 4^{n}$.  Hence, for finite $n$, Thm~\ref{eqn:catalysis_no_go} applies and we can conclude $\ket{\varphi}^{\otimes n} \nrightarrow_{P} \ket{H_{0}}$.  However, a supply of $\ket{\varphi}$ states is a resource for universal quantum computation.  This paradox is resolved by observing that although an exact $\ket{H_{0}}$ is impossible to produce, we may approximate $\ket{H_{0}}$ with a fidelity that asymptotically approaches unity as $n$ increases.  Similarly, in entanglement theory many copies of any pure entangled state may be converted into any other state in the asymptotic limit.

\section{}
\label{TwoQubitReductions}

We consider two qubit stabilizer measurements on the state $\sigma(q)$, defined in Thm.~\ref{Thm_Activation}, which is an incoherent mixture of $\ket{T_{0,1}}$ and $\ket{T_{1,0}}$.   We shall exploit that the state is invariant under $T$ rotations of either qubit A or B, such that
\begin{equation}
	T_{A}^{a}T_{B}^{b} \sigma(q) ( T_{A}^{a}T_{B}^{b} )^{\dagger} =  \sigma(q)
\end{equation}
for any integers $a$ and $b$.  Furthermore, for any Pauli operator $P_{A} P_{B}$, where $P_{A, B}=\{ X_{A, B}, Y_{A, B}, Z_{A, B} \}$, there exists integer $a$ and $b$ such that:
\begin{equation}
	T_{A}^{a}T_{B}^{b} P_{A}P_{B} ( T_{A}^{a}T_{B}^{b} )^{\dagger} = Z_{A}Z_{B}.
\end{equation}
Combing these two properties of the $T$ rotation, we have that for any two-qubit Pauli projection
\begin{equation}
\begin{array}{l}
	T_{A}^{a}T_{B}^{b}    (  \unity \pm P_{A}P_{B}) \sigma(q) 	  (  \unity \pm P_{A}P_{B})   ( T_{A}^{a}T_{B}^{b} )^{\dagger}   \\
	= 	(  \unity \pm Z_{A} Z_{B}) \sigma(q) 	  (  \unity \pm Z_{A}Z_{B}) .
\end{array}
\end{equation}
This symmetry entails that we only have to consider two possible Pauli projections, $\Pi_{\pm}=(\unity \pm Z_{A}Z_{B})/2$. As an intermediate step in our proof, we see that when the state is pure, $q=1$, two $T$-states can be probabilistically converted into a pure state on the Bloch sphere equator. This equatorization is an important step in using these resource for implementing non-Clifford gates.

First, we note that $T$-states in the computational basis are
\begin{eqnarray}
	\ket{T_{0} } & = & \cos ( \beta ) \ket{ 0 } + e^{i \pi / 4}  \sin ( \beta ) \ket{ 1 } , \nonumber \\
	\ket{T_{1} } & = & \sin ( \beta ) \ket{ 0 } - e^{i \pi / 4}  \cos ( \beta ) \ket{ 1 } ,
\end{eqnarray}
where $ \cos( 2 \beta )=1/\sqrt{3}$.  If we consider the $\Pi_{+}$ projection onto the even parity subspace then
\begin{equation}
	\Pi_{+} \ket{T_{0,1} }=\Pi_{+} \ket{T_{1,0} }= \cos(\beta) \sin(\beta)(  \ket{0,0} - i \ket{1,1}  ) .
\end{equation}
Since either pure state is projected onto the same stabilizer state, so too is the mixture $\sigma(q)$.  

For the odd parity projector, $\Pi_{-}$, the analysis is more involved as
 \begin{eqnarray*}
	\Pi_{-} \ket{T_{0,1} } & = & e^{i \pi / 4} \left[   \sin^2(\beta) \ket{1,0} - \cos^2(\beta)  \ket{0,1} \right] , \\
	\Pi_{-} \ket{T_{1,0} } & = & e^{i \pi / 4} \left[  \sin^2(\beta) \ket{0,1} - \cos^2(\beta)  \ket{1,0} \right] , 
\end{eqnarray*}
which are distinct nonstabilizer states.  Using the decoding $\ket{0,1} \rightarrow \ket{-}$ and $\ket{1,0} \rightarrow -i \ket{+}$, these states map to points on Bloch sphere equator,
 \begin{eqnarray*}
	\Pi_{-} \ket{T_{0,1} } & \rightarrow & \ket{\gamma_{+}}= (\ket{0}+e^{i \gamma} \ket{1} ) /\sqrt{2}, \\
	\Pi_{-} \ket{T_{1,0} } & \rightarrow & \ket{\gamma_{-}}= (\ket{0}+e^{-i \gamma} \ket{1} ) /\sqrt{2}, 
\end{eqnarray*}
where $\gamma=\pi/6$.  Since $\ket{\gamma_{+}}$ is in the positive octant of the Bloch sphere, no other decoding gets closer to the target $\ket{T_{0}}$ state.  Applying this analysis to the projection of the initial mixed state gives
\begin{equation}
	\Pi_{-} \sigma(q) \Pi_{-} \rightarrow q  \kb{\gamma_{+}}{\gamma_{+}} + (1-q)  \kb{\gamma_{-}}{\gamma_{-}} .
\end{equation}
The fidelity of this output with respect to $\ket{T_{0}}$ is
\begin{equation}
	f''' = q | \bk{T_{0}}{\gamma_{+}} |^{2} +(1-q) | \bk{T_{0}}{\gamma_{-}} |^{2} ,
\end{equation}
where,
\begin{equation}
	 | \bk{T_{0}}{\gamma_{\pm}} |^{2} = (9 \pm \sqrt{3})/12 .
\end{equation}
Comparing the fidelity  $f'''$ with $f''$ of Eq.~(\ref{1qubitReductFID}), we find that $f'''$ is always smaller.  Hence, no two qubit stabilizer projections can outperform the single qubit projection.

\section{}
\label{App:graphcode}

This appendix provides a proof of lemma~\ref{lem:graphcode}, which gives the maximum probability of projection onto a $2^{m}$-dimensional stabilizer-subspace.  All stabilizer subspaces are local-Clifford equivalent to a linear graph-code~\cite{GraphCodes1}, such that
\begin{equation}
	\Pi =  C _{\mathrm{loc}} \Pi_{G} C _{\mathrm{loc}}^{\dagger}.
\end{equation}
Our proof utilizes this local equivalence, so first we give a brief account of graph codes and their relevant features.   A graph code is defined by a graph $G$ and a $m$-dimensional linear code $\mathcal{C}$ over $\mathbb{Z}_{2}$.  We use $\ket{G}$ denote the graph state corresponding to graph $G$, which has stabilizer generators
\begin{equation}
	k_{j} = X_{j} \bigotimes_{k \in N(j)} Z_{k} , 
\end{equation}
where $N(j)$ denotes the set of vertices in the graph connected to vertex $j$.   The subspace for the graph-code is spanned by orthogonal graph states
\begin{equation}
	\ket{G_{\vec{c}}} =  Z_{\vec{c}} \ket{G} ,
\end{equation}
where $\vec{c}$ are binary vectors in the code $\mathcal{C}$, and $Z_{\vec{c}}=\bigotimes Z_{j}^{c_{j}}$.  The projector onto the graph code subspace is then
\begin{equation}
	\Pi_{G} = \sum_{\vec{c} \in \mathcal{C}} \kb{G_{\vec{c}}}{G_{\vec{c}}}.
\end{equation}
For our purposes we need to express this projector in terms of the graph code stabilizer $\mathcal{S}$ 
\begin{equation}
	\Pi_{G} = \frac{1}{2^{n-m}} \sum_{s \in \mathcal{S}} s .
\end{equation}
The stabilizer of the graph code is
\begin{equation}
\label{canonical}
	\mathcal{S} \equiv \{ s_{\vec{y}} = \prod_{j} k_{j}^{y_{j}} \vert \vec{y} \in  \mathcal{C}^{\perp} \} ,
\end{equation}
where $\mathcal{C}^{\perp}$ is the dual of $\mathcal{C}$.  Allowing for local Clifford unitary operators, the stabilizer of the graph code is $\mathcal{S}'=C_{\mathrm{loc}}\mathcal{S} C_{\mathrm{loc}}^{\dagger}$, and so the actual projector is
\begin{equation}
	\Pi = \frac{1}{2^{n-m}} \sum_{s' \in \mathcal{S}'} s'.
\end{equation}
In our proof we use the following fact:  The stabilizers of $\mathcal{S}'$ have the same weights as the those in the locally equivalent graph code stabilizer, $\mathcal{S}$.  That is, if $w(s)$ is weight of $s$  (the number of nonidentity Pauli operators) then $w(s')=w(C_{\mathrm{loc}}sC_{\mathrm{loc}}^{\dagger})=w(s)$, which holds because local Cliffords conjugate Pauli operators without changing their weight.  We use this fact in combination with other features of graph codes.

As for the relevant quantum state, this also has a Pauli decomposition.  Using that a single $T$-magic state is
\begin{equation} 
	\tau_{0} = \frac{1}{2} \left( 1 + \frac{X+Y+Z}{\sqrt{3}} \right),
\end{equation}
it follows that $n$ copies may be represented as
\begin{equation}
	\tau_{0}^{\otimes{n}} = \frac{1}{2^{n}} \sum_{g \in \mathcal{G} } g  \left( \frac{1}{\sqrt{3}} \right)^{w(g)} ,\
\end{equation}
where $\mathcal {G}$ is the set of Pauli operators with positive phase.  Hence, the projection probability is
\begin{eqnarray}
	\frac{\tr (  \Pi \tau_{0}^{\otimes n}  )}{2^{m-2n}} & = & \tr \left[   \sum_{s' \in \mathcal{S}' , g \in \mathcal{G}} s' . g  \left( \frac{1}{\sqrt{3}} \right)^{w(g)}   \right] , \\ \nonumber
	& = &    \sum_{s' \in \mathcal{S}' , g \in \mathcal{G}} \tr \left( s' . g \right)  \left( \frac{1}{\sqrt{3}} \right)^{w(g)}   .
\end{eqnarray}
The trace vanishes except when $g.s'=\pm \unity$, and so
\begin{equation}
	\frac{\tr (  \Pi \tau_{0}^{\otimes n}  )}{2^{m-n}} = \sum_{s' \in \mathcal{S}'  } \sign(s')  \left( \frac{1}{\sqrt{3}} \right)^{w(s')} ,
\end{equation}
where $\sign(s')$ is $\pm 1$, matching the phase of $s'$.  Clearly, an upperbound is established when all signs are positive, and hence
\begin{equation}
\label{EQN:now_graph}
	\frac{\tr (  \Pi \tau_{0}^{\otimes n}  )}{2^{m-n}} \leq   \sum_{s' \in \mathcal{S}'  } \left( \frac{1}{\sqrt{3}} \right)^{w(s')} .
\end{equation}
Having arrived at an inequality purely dependent on the weights of $\mathcal{S}'$, we can use $w(s')=w(s)$, to switch to the locally equivalent graph code. To determine the graph code weights $w(s)$ we use the decomposition in terms of canonical generators expressed in Eq.~(\ref{canonical}), where every $s_{\vec{y}}$ is identified with a binary vector $\vec{y} \in \mathcal{C}^{\perp}$,
\begin{equation}
	w(s_{\vec{y}})=w\left(  \prod_{j} k_{j}^{y_{j}}  \right) .
\end{equation}
When multiplying generators together the $X_{j}$ contributions can change into $\pm Y_{j}$, but never reduce in weight.  Hence,
\begin{equation}
	w(s_{\vec{y}}) \geq w\left( \vec{y} \right) ,
\end{equation}
where the R.H.S. is the weight, number of 1 entries, in the bit string $\vec{y}$.  Combining this result with Eq.~(\ref{EQN:now_graph}), we have
\begin{equation}
\label{EQN:now_graph2}
	\frac{\tr (  \Pi \tau_{0}^{\otimes n}  )}{2^{m-n}} \leq   \sum_{\vec{y} \in \mathcal{C}^{\perp}  } \left( \frac{1}{\sqrt{3}} \right)^{w(\vec{y})} .
\end{equation}
This inequality now depends solely on the classical linear code $\mathcal{C}^{\perp}$.  All such linear codes can, up-to relabeling of bits, be diagonalized such that the generator matrix, $M$, has an identity over the first $n-m$ elements, such that $M=[ \unity_{n-m} , M' ]$.  Dividing the bit strings into two halves $\vec{y}=( \vec{y}', \vec{y}'' )=(y'_{1},..,y'_{n-m}, y''_{1},..., y''_{m} )$, then the elements of $\vec{y}'$ are fixed by diagonlization of the generator matrix.    Furthermore, since $w(\vec{y})=w(\vec{y}')+w(\vec{y}'')$, we can conclude
\begin{eqnarray}
\tr (  \Pi \tau_{0}^{\otimes n}  ) & \leq & 2^{m-n}   \sum_{\vec{y} \in \{0,1\}^{n-m} }  \left( \frac{1}{\sqrt{3}}\right)^{w(\vec{y}')+w(\vec{y}'')}  , \\ \nonumber
\end{eqnarray}
The weights of $w(\vec{y}')$ are fixed by diagonalization, but $w(\vec{y}'')$ depend on features of the code.    We are interested in an upperbound, maximized over all possible projectors $\Pi$, which can be achieved when $w(\vec{y}'')=0$, and so
\begin{eqnarray}
\max_{\Pi} \left[ \tr (  \Pi \tau_{0}^{\otimes n}  ) \right] & \leq & 2^{m-n}   \sum_{\vec{y} \in \{0,1\}^{n-m} }  \left( \frac{1}{\sqrt{3}}\right)^{w(\vec{y}')}  , \\ \nonumber
& = &  2^{m-n}  \left(1+\frac{1}{\sqrt{3}} \right)^{n-m} = f_{\mathrm{st}}^{n-m}
\end{eqnarray}
This gives an upperbound, but it is easy to verify that it is saturated by measuring $m$ qubits in the computational basis and postselecting on ``+1" outcomes.

\section{}
\label{APP:Correlated_Noise_Resources}

In our magic activation protocol we saw that $\sigma(q)$ states (defined in Theorem~\ref{Thm_Activation}) can be a powerful resource for activation of bound families of states.   This suggests that they may be a powerful resource in their own right, and that the ability to prepare many copies of them may enable universal quantum computation.  This is the problem we address here, although for a more general class of states.  We consider states of the form
\begin{eqnarray}
\label{eqn_CNR}
\sigma( q, r  ) & = & q \tau_{1,0} + (1-q-2r) \tau_{0,1} \nonumber \\
 & + & r \left( \tau_{0,0} + \tau_{1,1} \right) ,
\end{eqnarray}
Before continuing let us reflect on some properties of these states.  Firstly,  states in this class are always separable.  Secondly, they are correlated states except for specific values $r=\sqrt{q}(1-\sqrt{q})$ which give a product state. Although this class of states is not completely general, any state can, by a suitable twirling procedure (See Appendix~\ref{Twirling}) be brought into this form.  

\begin{figure}
\centering
\includegraphics{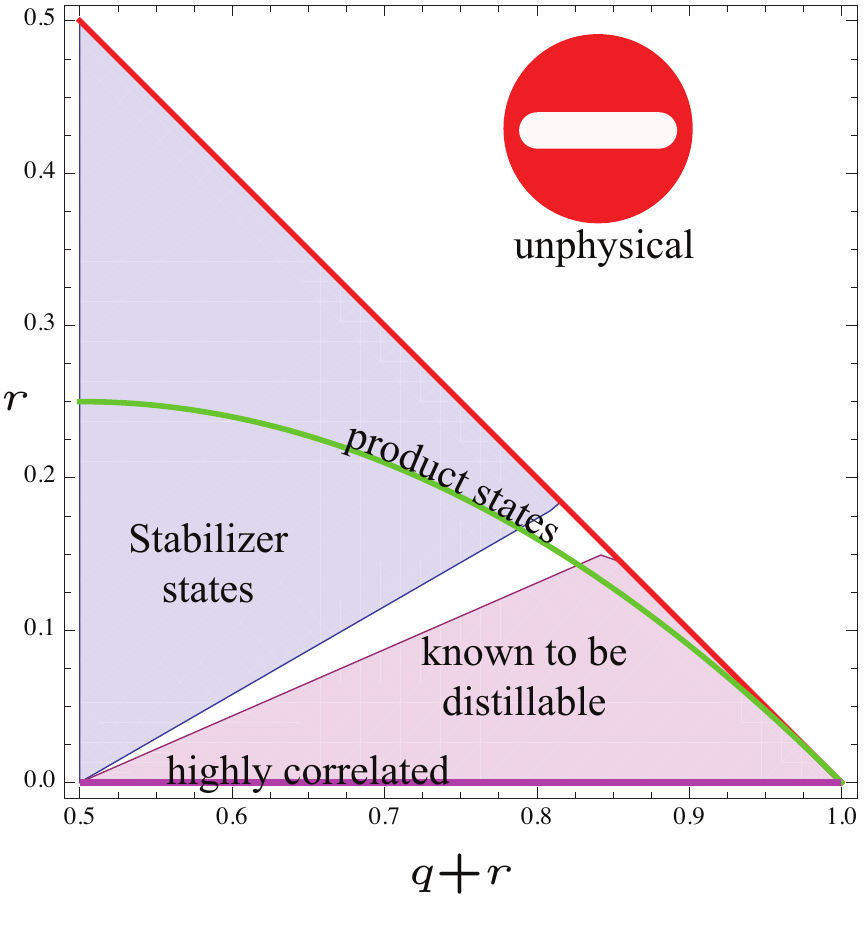}
\caption{The ``phase" diagram for correlated noise resource, $\sigma(q,r)$, as described by Eq.~(\ref{eqn_CNR}).  The diagram shows the region of stabilizer states, and the region of resources that are known to be universal for quantum computing.  Universality is possible if the daisy chain protocol achieves a fidelity, see Eq.~(\ref{Eqn:DaisyChainFid}), that exceeds the threshold above which the 5-qubit code can be utilized as in Ref.~\cite{BraKit05}.}
\label{fig:CorrelatedPhases}
\end{figure}

We now outline a protocol for exploiting many copies of these correlated states, where the maximum achievable fidelity approaches 1 as $r \rightarrow 0$.  Since the protocol consists in a chain of projections, we refer to the protocol as the \textit{daisy chain} protocol
\begin{enumerate}
	\item Prepare $n$ copies of the state $\sigma( q, r  )$, with the first pair as qubits $A$ \& $B$, pair 2 as qubits $C$ \& $D$, and so on;
	\item Measure qubit $A$ in the computational basis, and postselect on a ``$+1$" outcome;
	\item For qubits $B$ \& $C$, measure $X_{B}X_{C}$ and $Z_{B}Z_{C}$, and postselect on ``-1" outcomes for both;
	\item Perform the preceding step for qubits $D$ \& $E$, and all subsequent pairs;
	\item Leave the final qubit unmeasured and discard all measured qubits.
\end{enumerate}
After step 2, qubit $B$ is left in a product state $\tau(f_{0})$ with fidelity
\begin{equation}
	f_{0} = \frac{q (1 -f_{\mathrm{st}})  + r f_{\mathrm{st}}}{r + q (1-f_{\mathrm{st}})   +(1-q-2r)f_{\mathrm{st}}},
\end{equation}
which is a generalization of Eq.~(\ref{1qubitReductFID}).

Step 3 results in the familiar singlet projection on the second and third qubits ($B$ and $C$).  After this projection, qubit $D$ is in the state $\tau(f_{1})$, where $f_{1}$ is determined by the matrix equation
\begin{eqnarray}
\left( \begin{array}{c} f_{1} \\ 1-f_{1} \end{array} \right) & \propto & \left(   \begin{array}{cc} q & r \\ r & (1-2r-q) \end{array}  \right) . \left( \begin{array}{c} f_{0} \\ 1-f_{0} \end{array} \right),
\end{eqnarray}
where we use a proportionally sign as the left most vector must be renormalized to obtain the fidelity.  This is then repeated between every $(2j+1)^{\mathrm{th}}$ and $(2j+2)^{\mathrm{th}}$ qubit.  After all $n-1$ singlet projections, we find that the last qubit is left in the state $\tau(f_{n-1})$, where
\begin{eqnarray*}
\left( \begin{array}{c} f_{n-1} \\ 1-f_{n-1} \end{array} \right) & \propto & \left(   \begin{array}{cc} q & r \\ r & (1-2r-q) \end{array}  \right)^{n-1}  . \left( \begin{array}{c} f_{0} \\ 1-f_{0} \end{array} \right).
\end{eqnarray*}
The limiting behavior, for large $n$, of this matrix equation is determined by the matrix eigenvalues,  $\mu_{1}$ and $\mu_{2}$, and eigenvectors.  Whenever eigenvalues have different magnitudes, the matrix (as $n \rightarrow \infty$) projects onto the eigenvector~\footnote{Note that in this context a eigenvector corresponds to a particular $(f, 1-f )$, and hence specifies a mixed state.} with the largest eigenvalue.   As one would expect, when $\sigma(q,r)$ is a product state the eigenvalues are identical, but in all other cases there is one dominant eigenvalue which determines a limiting fidelity
\begin{equation}
\label{Eqn:DaisyChainFid}
	\lim_{n \rightarrow \infty}  f_{n} = \left\{   1 + \tan \left[   \frac{1}{2} \arctan \left( \frac{2r}{2(r+q)-1} \right)  \right]  \right\}^{-1} .
\end{equation}
Convergence to this fidelity is exponentially fast in the number, $n$, of copies of $\sigma(q,r)$.  Specifically, for large but finite $n$, deviations from this fidelity vanish as $(\mu_{2}/\mu_{1})^{n}$, where $\mu_{2}$ is the smaller eigenvalue.  In Fig.~\ref{fig:CorrelatedPhases} we chart out various parameter regimes indicating when the resource is a stabilizer state and when it provides a resource for universal quantum computing.  Universality may be achieved by a combination of the daisy chain protocol followed by the standard 5-qubit distillation procedure~\cite{BraKit05}.  

Notice that product states are not the only nonstabilizer states where we observe a regime where no known methods enable universal quantum computing.  There is a temptation to conjecture that some notion of boundness applies to any family of states that transverses the gap anywhere except via the origin.  However, the daisy chain protocol shows that all nonproduct states can be purified towards some state, even if that state is not above the threshold for the 5-qubit code.  As such our current definition for boundness would not extent to these families.   However, this seems like a failing of our definition more than anything else, as the extent of possible purification appears to be limited by how correlated the raw resource is.

Finally, note that a very similar protocol and analysis can be performed for 2-qubit correlated states in any basis, not just the $T$-basis.

\section{}
\label{Twirling}

Here we outline twirling protocols for bringing an arbitrary state into the form $\sigma( q, r )$, as defined in Eq.~(\ref{eqn_CNR}). We perform the following:
\begin{enumerate}
\item randomly choose a unitary operator from the set $\{ 1, T, T^{2} \}$ and apply to qubit A;
\item randomly choose a unitary operator from the set $\{ 1, T, T^{2} \}$ and apply to qubit B;
\item randomly choose a unitary operator from the set $\{ 1, Y_{A}H_{A} SWAP_{A,B} H_{A} Y_{A} \} $ and apply.
\end{enumerate}
The first two steps diagonalize the state in the $\ket{T_{i,j}}$ basis, and the third step mixes the symmetric terms.

\end{document}